%% file: main.tex
\documentclass{article}

\usepackage{PRIMEarxiv}

\usepackage[utf8]{inputenc} 
\usepackage[T1]{fontenc}    
\usepackage{hyperref}       
\usepackage{url}            
\usepackage{booktabs}       
\usepackage{amsfonts}       
\usepackage{nicefrac}       
\usepackage{microtype}      
\usepackage{lipsum}
\usepackage{fancyhdr}       
\usepackage{graphicx}       
\usepackage{natbib}
\usepackage{amsmath}
\usepackage{multirow}
\graphicspath{{media/}}     

\newcommand{\eg}{\emph{e.g.},\ }

\title{Wave climate on the southwestern coast of Lake Michigan: Perspectives from wave directionality
\thanks{\textit{\underline{Citation}}: 
\textbf{Lu, B., Wang, W., Wu, C. and Liu, Y., 2026. Wave climate on the southwestern coast of Lake Michigan: Perspectives from wave directionality. Ocean Engineering, 343, p.123306.DOI:https://doi.org/10.1016/j.oceaneng.2025.123306} 
}}

\author{
Boyuan Lu \\
Department of Civil and Environmental Engineering \\
University of Wisconsin, Madison \\
Madison\\
  \texttt{blu38@wisc.edu} \\
   \And
   Wei Wang \\
   Department of Civil and Environmental Engineering \\
University of Wisconsin, Madison \\
Madison\\
  \texttt{wwang487@wisc.edu} \\
  \And
Chin Wu \\
   Department of Civil and Environmental Engineering \\
University of Wisconsin, Madison \\
Madison\\
  \texttt{chin.wu@wisc.edu} \\
   \And
Yuli Liu \\
  School of Marine Sciences  \\
Nanjing University of Information Science\\
and Technology  \\
Nanjing\\
  \texttt{liu.yuli@nuist.edu.cn } \\
}

\begin{document}
\maketitle

\begin{abstract}
Wave directionality plays a critical role in shaping coastal conditions and
influencing local livelihoods, underscoring the importance of conducting
detailed analyses. This study examines directional wave climate along the
southwestern coast of Lake Michigan from 1979 to 2023 using the Directional Wave
Entropy (DWE). Directionality was characterized in terms of inter-annual trends,
monthly patterns, spatial variation, and extreme wave conditions. Overall,
results exhibited a strong bi-directionality, with dominant northern and
southern wave systems along the coast of our study site. A few annual trends for
the inter-annual wave climate were observed, and there is a clear seasonal
variation such that bi-directionality increases in the summer and winter
seasons. As for spatial variation of wave directionality, all locations in the
study sites presented a bi-directional wave climate. The two dominant directions
of wave directionality: northern and southern mean significant wave heights were
also characterized in all locations of study sites as 0.566 and 0.563 meters.
Furthermore, the extreme wave heights in the northern direction are
significantly greater than the extreme waves in the southern direction. In
summary, these findings suggest the importance of wave directionality on coastal
structural design and coastal morphology management along the coast of our study
site. 
\end{abstract}

\keywords{Wave climate  \and Wave directionality \and Inter-annual trend \and Monthly trend \and Extreme wave conditions \and  Lake Michigan}

\section{Introduction}
\label{c3_Introduction}

Wind-waves are important processes that significantly affect the evolution of
coastal morphology, coastal structural design, and coastal ecosystem health
\citep{casas2024wind}. To describe features of wind-waves, the wave climate,
which is defined by average and extreme wave conditions such as wave height,
period, energy, and directionality,has been widely applied across engineering
and scientific fields 
\citep{wiegel_oceanographical_1964,wiegel_oceanographical_2013}. First, wave
height and direction are known to be associated with nearshore sediment
transport \citep{pethick_introduction_1984,davila_promoting_2014}, which can
imply the changes in coastal morphology
\citep{lamoe_wave_1989,benumof_relationship_2000,brown_factors_2005}. Second,
wave climate is crucial for designing coastal structures. For example, wave
height and period are key parameters for designing harbors, marinas, and
breakwaters to shelter ships and mitigate harbor resonance
\citep{rabinovich_seiches_2009}.
Additionally, extreme wave conditions are fundamental considerations in the
development of wave power facilities for renewable energy 
\citep{guillou_wave_2020,neary_characterization_2020}. Third, wave climate is
associated with the health condition of nearshore ecosystems. For example, wave
energy was found to correlate with water quality metrics, such as water
bacterial exceedances, in beach environments \citep{feng_wave_2016}. Meanwhile,
wave energy can influence the oxygen dissolution of water and further affect the
spawning and egg-dispersal of trout
\citep{sly_interstitial_1988,fitzsimons_relationship_2014}. In summary, wave
climate is a critical concept that facilitates the understanding of wave impacts
on coastal environments.

Directional wave climate is a special perspective of wave climate, which focuses
on the directional characteristics of the waves. Unlike the conventional “wave
directionality” which reflects the feature of the instant energy spectrum, the
directionality of wave climate describes the directional features over time
\citep{wiggins_coastal_2019,wiggins_regionally-coherent_2019}. In most open
oceanic areas, the waves typically approach from a uniform direction
\citep{echevarria_seasonal_2019, echevarria_influence_2020}, which is also known
as a uni-directional wave climate. Nevertheless, in some specific regions, waves
can be bi-directional or multi-directional, where waves come from more than one
dominant wave direction. For example, wave bi-directionality has been found in
enclosed or semi-sheltered water bodies with elliptical shapes such as Black Sea
\citep{gippius_black_2020}, Azov Sea \citep{amarouche_wind-sea_2024}, Baltic Sea
\citep{mannikus_directional_2023}, Bohai Sea \citep{miao_study_2024}, Adriatic
Sea \citep{mannikus_directional_2023}, and Red Sea \citep{shamji_extreme_2020}.
Bi-directional wave climate can also occur at the channels, such as English
Channel \citep{wiggins_regionally-coherent_2019}, Tasman Sea
\citep{mortlock_directional_2015}, and Malacca Strait
\citep{aboobacker_wave_2017}, where the wave could approach from two sides.
Given the widespread presence of bi-directional wave climates across the globe,
a thorough characterization and analysis of their features and implications is
essential.

Characterizing wave directionality is imperative because different
directionalities can have different implications on coastal environments such as
coastal morphology \citep{ranasinghe_southern_2004}. Specifically, a
uni-directional wave is likely to induce a uniform longshore sediment transport
\citep{scott_role_2021}, leading to beach erosion at the updrift extent
\citep{klein_short-term_2002}. In contrast, a bi-directional wave can induce
longshore sediment transport towards opposite directions, which is likely to
cause an imbalance of the sediment budget in the coastal littoral cells
. Imbalance of sediment budget in a littoral cell is
usually a signal of shoreline erosion and accretion
\citep{hapke_review_2010,lopez-olmedilla_effect_2022}. Besides beach morphology,
wave directionality can also affect coastal structural designs. The conventional
design wave height for designing structures assumes that the waves are
uni-directional \citep{sanil_kumar_design_2004}. Nevertheless, this parameter
can be underestimated when considering the distribution of two components of the
bi-directional waves \citep{sanil_kumar_design_2004}, leading to instability of
structures. Last but not least, bi-directional waves can generate larger ship
responses than uni-directional waves, leading to navigation safety issues
\citep{huang_cfd_2021}. To date, wave directionality characterization is crucial
for coastal morphology, structures, and ships.

Despite that wave directionality is important and has been characterized in some
specific oceanic areas, the wave directionality has rarely been fully explored
in freshwater lake systems. Lake Michigan, the third largest lake among the
Great Lakes, is significantly affected by wind waves
\citep{huang_impacts_2021,huang_wave_2021}. Previous studies have been made to
reveal the wave climate in Lake Michigan, mainly focusing on inter-annual trends
\citep{olsen_long_2019,jabbari_increases_2021}, monthly fluctuation
\citep{meadows_relationship_1997,huang_wave_2021}, spatial variabilities
\citep{huang_wave_2021}, and extreme conditions
\citep{sogut_characterizing_2018} of wave height. For directionality, some
locations in Lake Michigan typically experience waves from both northern and
southern directions due to their extensive north-south wind fetch, as reported
in several studies 
\citep[\eg][]{davidson-arnott_wave_1980,booth_wave_1994,olsen_long_2019,abdelhady_shoreline_2025}.
Nevertheless, two knowledge gaps remain. First, the current wave directionality
study in Lake Michigan is not based on a quantitative approach. Second, the
spatial and temporal trends of the two dominant wave components of
bi-directional waves have not been adequately investigated.

The objective of this research is to characterize the directionality of wave
climate in the southeastern Wisconsin coast of Lake Michigan. An index-based
metric is developed to characterize wave directionality using the WIS hindcast
data from 1979 to 2023. Spatial patterns and temporal (\eg inter- and
intra-annual) trends of the directional wave climate at the study site were
examined. Wave directionality under extreme wave conditions (\eg 90\%, 99\%, and
100-yr return period) was revealed. This paper is structured as follows: Section
\ref{c3_Methods} introduces the study site and describes methods for
characterizing wave directionality, spatial and temporal features, and extreme
conditions. Section \ref{c3_Results} presents results. In Section
\ref{c3_Discussions} discusses several topics, including the sensitivity of the
threshold in defining wave directionality, the correlations of wave
directionality with wave and wind factors, wave directionality characterization
results for the whole Lake Michigan shoreline, implications of wave
directionality on beach morphology, and limitations of the study. Section
\ref{c3_Conclusion} provides conclusions.

\section{Methods}
\label{c3_Methods}

\subsection{Study site and data sources}
\label{c3_Study site and data sources}

\begin{figure}[htbp]
  \centering
  \includegraphics[width=0.8\textwidth]{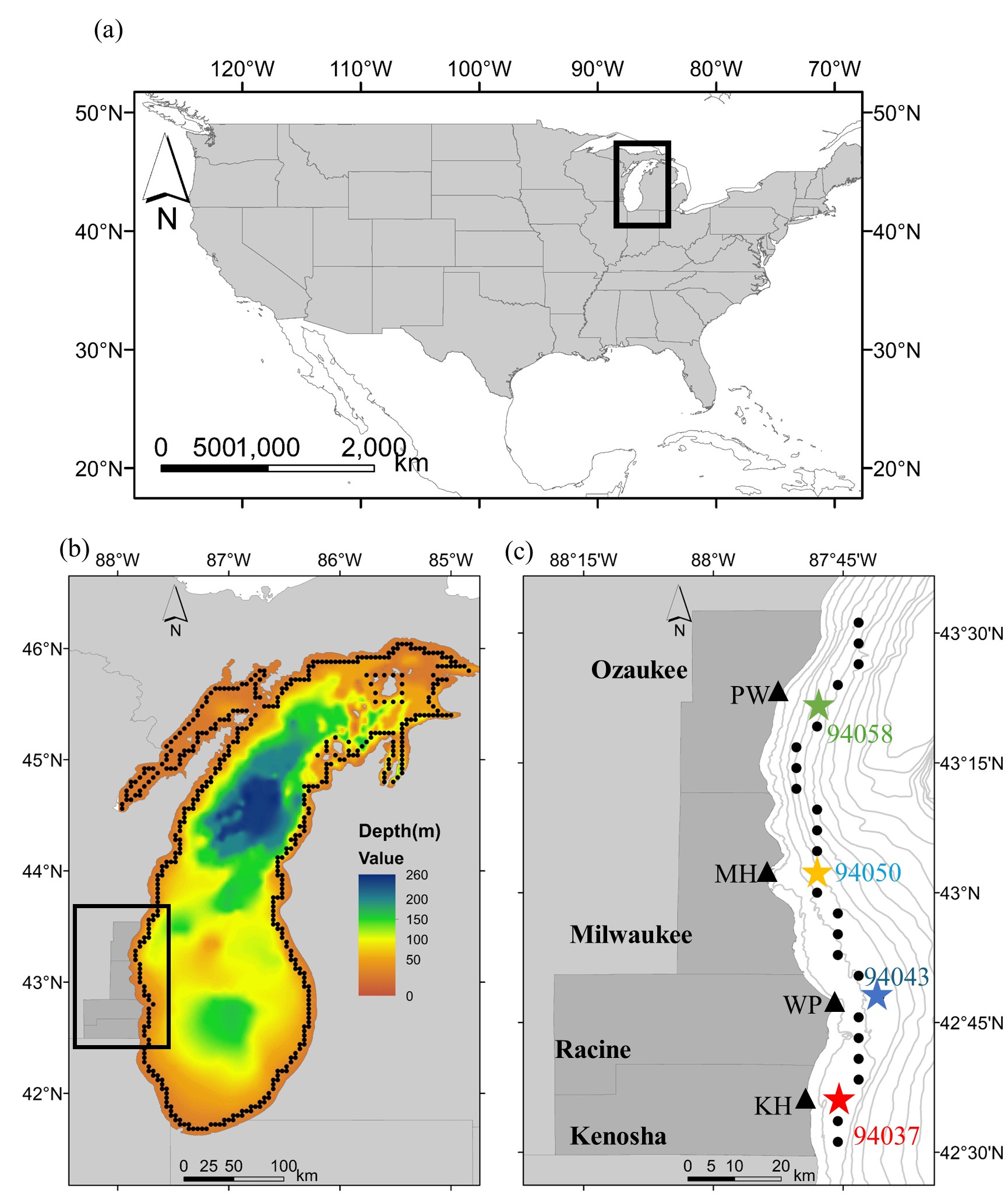}
  \caption{Study Sites. (a) the location of Lake Michigan in the United States. (b) Lake Michigan bathymetry and WIS stations. The 490 WIS stations are marked as dark dots, and the white diamond marks the deepest point of Lake Michigan. (c) study sites in Wisconsin area. Four major cities (PW-Port Washington, MIL-Milwaukee Harbor, WP-Wind Point, and KEN-Kenosha Harbor) are marked with dark triangles. The four stations for four cities are marked with colored stars, aligned with their corresponding station IDs. The interval of the bathymetry contour is 10 meters under IGLD1984. }
  \label{fig:fig3.1}
\end{figure}

The study area is the southwestern nearshore of Lake Michigan, USA (Figure
\ref{fig:fig3.1}a). Lake Michigan is one of the five Great Lakes of North
America and the only one located entirely within the United States. Figure
\ref{fig:fig3.1}b presents the geographical coverage and bathymetry of Lake
Michigan. The lake is surrounded by four states: Wisconsin, Illinois, Indiana,
and Michigan. Its deepest point, at 275 meters, is located in the northern part
(marked as a white diamond in Figure \ref{fig:fig3.1}b). The black dots
represent nearshore stations from the U.S. Army Corps of Engineers (USACE) Wave
Information Study (WIS), located 5–10 km offshore with water depths of 20–70 m.
Figure \ref{fig:fig3.1}c zooms into the gray area in Figure
\ref{fig:fig3.1}b, covering the region from the northern border of Ozaukee
County, Wisconsin, through Milwaukee County and Racine County, to the southern
border of Kenosha County. This is the most developed region in Wisconsin, with a
population exceeding 1.3 million and a GDP of 100 billion dollars (United States
Census Bureau, 2020). The study area spans approximately 120 kilometers (80
miles) of coastline, featuring bluffs, beaches, sandy dunes, and urbanized
areas. 

Long-term (44 years from 1979 to 2023) hourly wave data, including significant
wave height (Hs), peak spectral wave period (TP), and mean wave direction (MWD),
were obtained from the WIS \citep{hubertz_wind-waves_1991}. In the Great Lakes,
WIS provided hourly directional wave fields (72 bands, each 5 degrees) and
frequency ranges (28 bands starting at 0.0611 Hz with $f(n)=1.1f(n-1)\Delta f$,
with $f(n)$ representing frequency for band n) using the third-generation wave
model (WAM) (Komen et al., 1984). The WAM model simulated waves by 0.04-degree
spatially interpolated wind fields from the Climate Forecast System Reanalysis
(CFSR) \citep{saha_ncep_2010} and mean daily ice concentration
fields \citep{yang_consistent_2020}. A total of 26 WIS hindcast stations
(ST94035–ST94064) were used in this study, with four stations (ST94038, ST94045,
ST94059, ST94061) removed due to redundancy with inner stations. These 26
stations were marked as black dots or stars in Figure \ref{fig:fig3.1}c. Four
stations (94037, 94043, 94050, and 94058; star markers in Figure
\ref{fig:fig3.1}c) were selected to represent the local wave climate, as they
are located near four representative cities in the study area: PW (Port
Washington in Ozaukee County), MIL (Milwaukee City in Milwaukee County), WP
(Wind Point Village in Racine County), and KEN (Kenosha City in Kenosha County).

\subsection{Characterization of wave directionality}

\begin{figure}[htbp]
  \centering
  \includegraphics[width=0.8\textwidth]{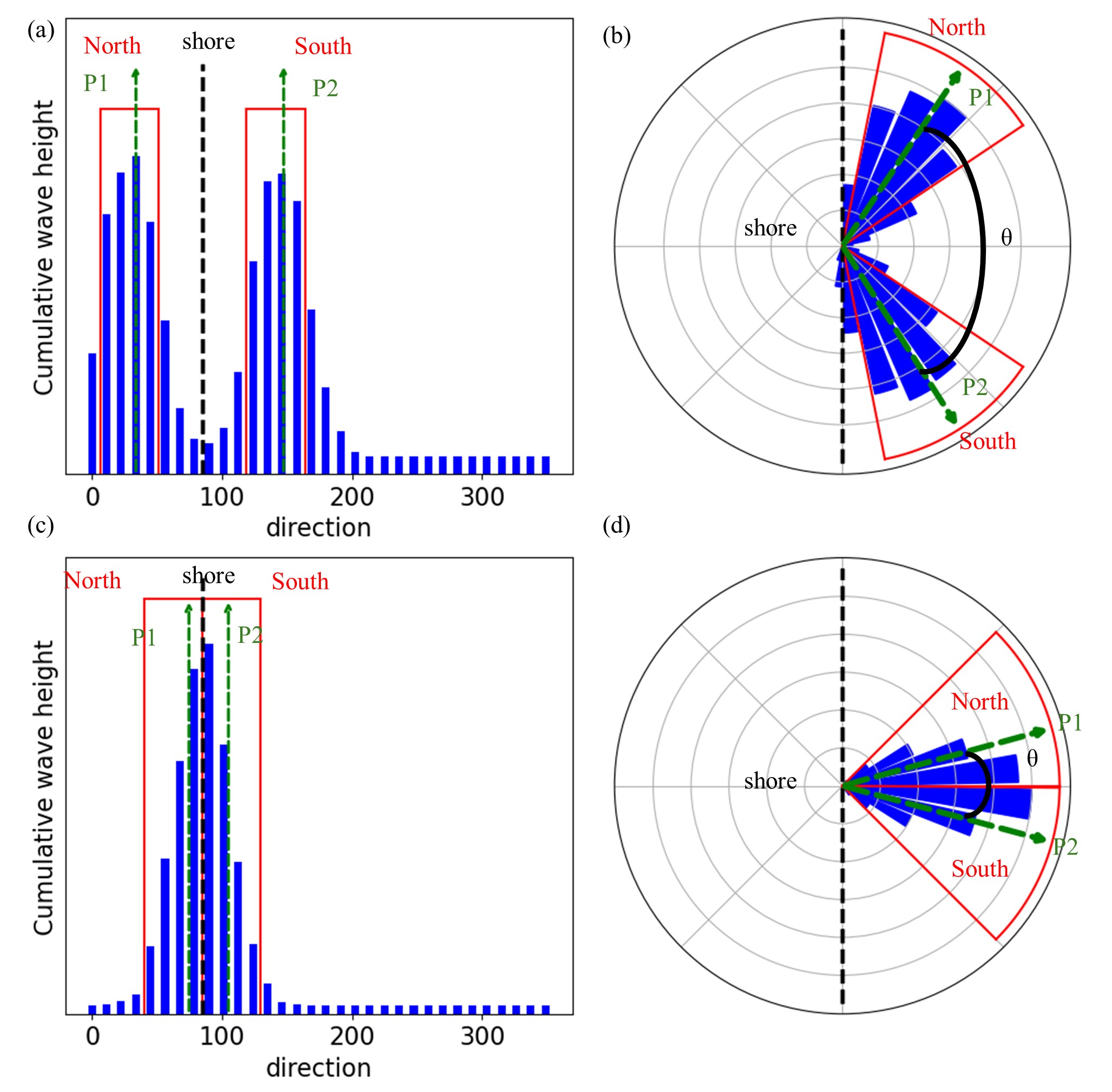}
  \caption{Characterization of two typical wave directionalities: uni- and bi-directionality. (a) and (b) are schematics of height-direction distribution and wave rose for bi-directionality, respectively. (c) and (d) are schematics of height-direction and wave rose for uni-directionality, respectively. The shoreline orientation is marked as a black dash line, while the characterizations of northern wave and southern waves are marked with red frames, respectively. The green arrows $P_1$ and $P_2$ represent the averaged waves for the northern and southern components. $\theta$ is the angle between these two averaged wave components.}
  \label{fig:fig3.2}
\end{figure}

Characterization of wave directionality involves both qualitative and
quantitative methods. From a qualitative perspective, wave data, including its
direction and height, was visualized using wave rose maps. For the wave rose,
incident wave angles were divided into 32 equally spaced spokes, and the wave
height was aggregated in 0.4-meter intervals. From a quantitative perspective,
wave directionality was characterized using indices, which summarized the
statistical features of wave climates among different wave components. To
capture the dominant wave components, wave instances are categorized into two
primary groups—northern and southern—based on wave directions relative to the
shoreline orientation, as illustrated in Figure \ref{fig:fig3.2}. Notably, wave
components with offshore-directed angles relative to the shoreline orientation
are filtered out, based on the assumption that most high-magnitude waves
approach from the inshore direction. The shoreline orientations at all WIS
stations were approximated by the nearby averaged shoreline orientation within
10 km. To further refine the dominant wave climate and reduce computational
complexity, the northern and southern dominant wave climates were identified by
averaging over four consecutive directional bins with the highest mean values,
each spanning 11.25 degrees. This is illustrated in Figure \ref{fig:fig3.2},
where the green arrows ($P_1$, $P_2$) represent the average wave heights, and
the red frames indicate the directional bins corresponding to the northern and
southern sectors. With the northern and southern wave components, an index-based
approach is applied to capture the relations among different wave components.
Several indices have been proposed in previous studies, such as the Wave
Directionality Index (WDI)
\citep{wiggins_coastal_2019,wiggins_regionally-coherent_2019} in Equation
\ref{eq:eq3.1},

\begin{equation}
    WDI = \frac{(PW_1-PW_2) - \overline{(PW_1-PW_2)}}{\sigma(PW_1-PW_2)}
\label{eq:eq3.1}
\end{equation}

where $(PW_1-PW_2)$ is the difference between the first and the second
directional wave power, $(PW_1-PW_2)$ is the long-term mean and $(PW_1-PW_2)$ is
the long-term standard deviation of that difference. Another approach is Wave
Bidirectionality Index (WBI) \citep{abdelhady_shoreline_2025} in Equation
\ref{eq:eq3.2},

\begin{equation}
    WBI = \frac{mean(PW) - abs(mean(PW_1)-mean(PW_2))}{mean(PW)}
\label{eq:eq3.2}
\end{equation}

where $PW$ is the total wave power at a certain site, $PW_1, PW_2$ follows the
same definition as the previous ones. Nevertheless, there is a limitation for
these two indices that these indices did not account for the angles between
dominant wave components. For example, if two wave groups—northern and
southern—originate from the northeast and southeast, respectively (a
bi-directional wave climate; Figure \ref{fig:fig3.2}a–b), the indices correctly
identify them as distinct components and properly capture the bi-directionality.
In contrast, if only one wave group exists (a uni-directional wave climate) with
angles perpendicular to the shoreline (Figure \ref{fig:fig3.2}c-d), the indices
may incorrectly split the eastern waves into two components, leading to a
misleading classification of the wave climate as bi-directional. To address this
issue, we developed an entropy-based index termed Directional Wave Entropy
(DWE). 

\begin{equation}
DWE = -\sum_{i,j}p_i\log_2(p_i)\sin{\frac{|\theta_i - \theta_j|}{2}}
\label{eq:eq3.3}
\end{equation}

where $p_i$ and $p_j$ represent frequency of two dominant waves weighted by
significant wave height and $\theta_i-\theta_j$ is the angle between two
dominant waves. Under an assumption of bi-directionality, the Equation
\ref{eq:eq3.3} can be simplified using averaged wave components $P_1,P_2$ as
below in Equation \ref{eq:eq3.4},

\begin{equation}
    DWE = - \frac{P_1}{P_1+P_2}\log_2\frac{P_1}{P_1+P_2}\sin{\frac{\theta}{2}}- \frac{P_2}{P_1+P_2}\log_2\frac{P_2}{P_1+P_2}\sin{\frac{\theta}{2}}
\label{eq:eq3.4}
\end{equation}

where $P_1, P_2$ indicates aggregation over all wave occurrences weighted by
significant wave height within the given period (1979–2023) and $\theta$ is the
angle between $P_1, P_2$. DWE is based on the concept of entropy, where greater
bi-directionality yields values closer to 1 and greater uni-directionality
yields values to zero. The inclusion of the $\sin{\frac{\theta}{2}}$ term in the
entropy formulation ensures a monotonic trend that the index approaches zero
when a uni-directional wave climate is oriented perpendicular to the shoreline
(e.g. the case in  Figure \ref{fig:fig3.2}c-d). After the indices were
computed, the directionality can be characterized using a predetermined
threshold –0.65. This threshold was chosen based on typical wave climate
characteristics, under the assumption that if the ratio of weighted occurrence
between two wave components exceeds two, or if the angular difference between
them is greater than 90 degrees, the wave climate can still be considered
bi-directional. A sensitivity test for this threshold was conducted in the
discussion.

\subsection{Inter-annual and intra-annual analysis}
\label{c3_Inter-annual and intra-annual analysis}

Inter-annual and intra-annual analyses for wave heights and DWE were conducted
to reflect the long-term characteristics of wave climate. For trend analysis,
hourly data were aggregated into annual subsets and monthly subsets using the
mean, top 90th percentile, and top 99th percentile wave heights for each WIS
site (PW: 94037, MIL: 94043, WP: 94050, and KEN: 94058). Since DWE is a derived
index and cannot be directly used to calculate percentiles, its 90th and 99th
percentiles were determined based on the cumulative wave height exceeding the
respective percentile threshold for 90-percentile and 0.99 for 99-percentile.
The Theil-Sen slope estimator was applied to each annual time series of wave
heights and DWE to determine the trend for mean, 90-percentile, and
99-percentile. This was followed by the Mann-Kendall test
\citep{kendall_new_1938,mann_nonparametric_1945}, which was conducted to assess
the statistical significance of the trend. The null hypothesis of no trend was
tested at a 95\% confidence level. To reflect the relative magnitude of
bi-directionality waves, the wave heights for northern and southern directions
(defined in Figure \ref{fig:fig3.2}) were extracted on the annual and monthly
subsets and were then subtracted with the total wave height from 0 to 359
degrees.

\subsection{Extreme value analysis}
\label{c3_Extreme value analysis}

The extreme wave analysis is intended to compute estimated return periods of
extreme wave events. To explore the directional patterns of extreme wave events,
the extremes analysis was conducted under the three subgroups proposed in
previous sections. In order to reveal the extreme wave events from an extensive
series of wave height data, a Peak-Over-Threshold (POT) method with a 95\%
quantile of significant wave height series as the threshold was employed. The
POT method calculated the cumulative probability by identifying sampled events
that exceeded a predefined threshold and then ranked these events by magnitude
to construct an empirical cumulative frequency distribution
\citep{bechle_meteotsunami_2015}. This approach enhanced analytical precision in
extreme event analysis \citep{coles_introduction_2001}. For sequences of wave
height data that persistently exceed the designated threshold, a clustering
methodology was applied, which treats all wave occurrences within a 48-hour
period as a singular, independent extreme wave event. The extreme wave events
extracted using the POT method were assumed to follow a Generalized Pareto
Distribution (GPD), as expressed in Equation \ref{eq:eq3.5}.

\begin{equation}
    F(x) =
\begin{cases}
1 - \left(1 + \dfrac{\xi x}{\sigma} \right)^{-\frac{1}{\xi}} & \text{if } \xi \ne 0 \\
1 - e^{-x/\sigma} & \text{if } \xi = 0
\end{cases}
\label{eq:eq3.5}
\end{equation}

where $x$ is the maximum wave height of an extreme wave event, $F(x)$ is the
cumulative probability over $x, \sigma$ is the scale factor, and $\xi$ is the
shape factor. In Equation \ref{eq:eq3.5}, the scale factor and shape factor were
estimated by Maximum Likelihood Estimation (MLE). 

\section{Results}
\label{c3_Results}

\subsection{Bi-directional wave directionality}
\label{c3_Bi-directional wave directionality}

\begin{figure}[htbp]
  \centering
  \includegraphics[width=0.8\textwidth]{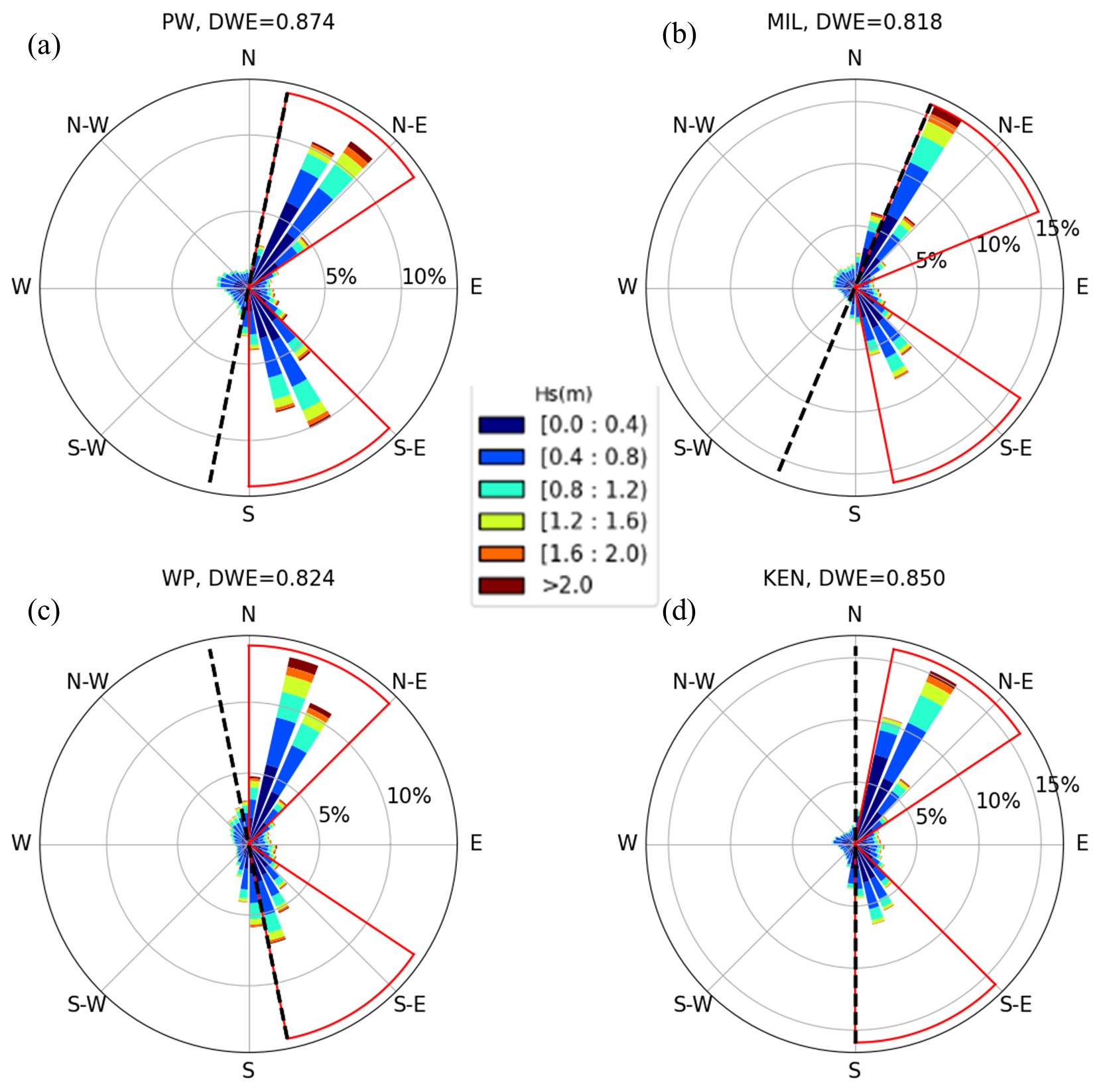}
  \caption{Wave rose maps and DWE at four selected locations. (a) PW-Port Washington, (b) MIL-Milwaukee Harbor, (c) WP-Wind Point, and (d) KEN-Kenosha Harbor with their corresponding directional wave entropy. The black dashed lines represent the shoreline orientation and red frames indicate the selected directional bins.}
  \label{fig:fig3.3}
\end{figure}

The wave directionality was characterized at nearshore WIS stations and
visualized using wave roses at four selected WIS stations from 1979 to 2023 in
Figure \ref{fig:fig3.3}. The wave roses in four locations presented a noticeable
bi-directional distribution in the NNE directions and SSE directions. Due to
different shoreline orientations at four sites (the black dash line in Figure
\ref{fig:fig3.3}), the north and south wave divisions (the red frame in Figure
\ref{fig:fig3.3}) were categorized differently among the four locations. At PW,
the northern wave division spanned from $11.25^\circ$ to $56.25^\circ$, while
the southern wave division ranged from $146.25^\circ$ to $191.25^\circ$. At MIL,
the northern division extended from $22.5^\circ$ to $67.5^\circ$, and the
southern from $157.5^\circ$ to $202.5^\circ$. At WP, the northern division
spanned from $-11.25^\circ$ to $33.75^\circ$, and the southern from
$123.75^\circ$ to $168.75^\circ$. At KEN, the northern division ranges from
$0^\circ$ to $45^\circ$, while the southern extends from $135^\circ$ to
$180^\circ$. The bi-directionality of the four sites was further confirmed by
their DWE values, which are 0.874, 0.818, 0.824, and 0.850. All four selected
WIS stations presented bi-directional wave climate, as their DWE values were
less than the defined threshold of 0.65. The wind roses and corresponding DWEs
were also computed and reported in Figure \ref{fig:fig3.1}a (see Chapter
Appendix) using the wind climate data from WIS. However, the applicability
of this approach to wind climate is limited, as discussed in the limitations
section.

\subsection{Inter-annual patterns}
\label{c3_Inter-annual patterns}

\begin{figure}[htbp]
  \centering
  \includegraphics[width=0.8\textwidth]{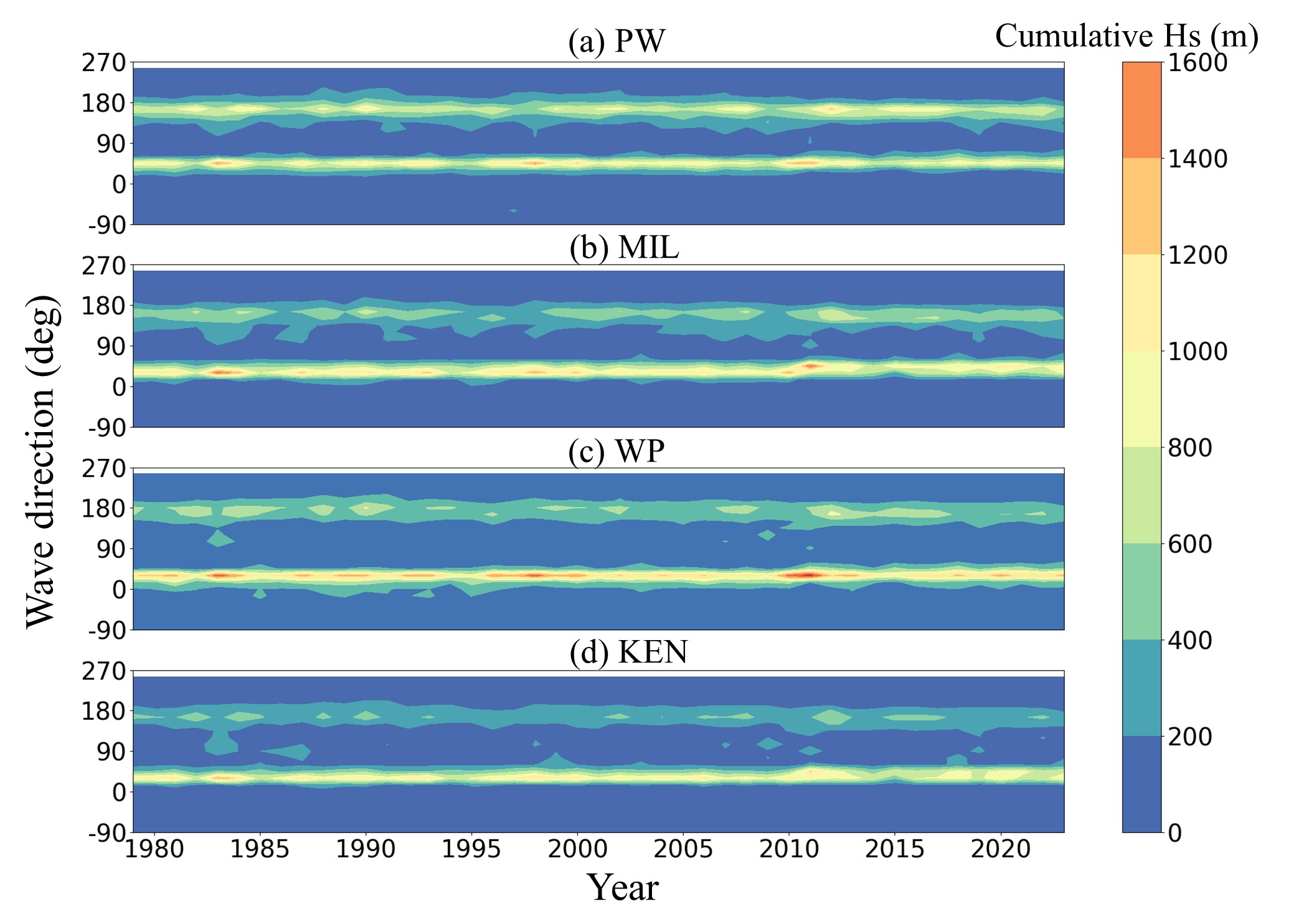}
  \caption{Directional-time wave height spectrum from 1979 to 2021 at the following locations : (a) PW-Port Washington, (b) MIL-Milwaukee Harbor, (c) WP-Wind Point, and (d) KEN-Kenosha Harbor.}
  \label{fig:fig3.4}
\end{figure}

Cumulative wave height spectra presented in Figure \ref{fig:fig3.4} provided
valuable insights into both the overall wave directionality and its temporal
variations. As shown, warmer colors represent higher cumulative wave heights
recorded annually for a specific direction, calculated at 15-degree intervals.
All four locations exhibited two distinct directional bands for most of the
examined years, indicating a consistent bi-directional wave climate over the 42
years. The lower bands (northern wave) for four sites were located in the
directional range between 15 and 45 degrees, while the higher bands (southern
wave) were located between 150 and 180 degrees. Additionally, the northern band
in the spectrum exhibits larger extreme wave heights than the lower band, as
there were warmer colors scattered in the higher band. This shows that extreme
wave heights were larger in the northern wave direction. Further analysis of
these extreme wave height characteristics is presented in the subsequent
sessions.

\begin{figure}[htbp]
  \centering
  \includegraphics[width=0.8\textwidth]{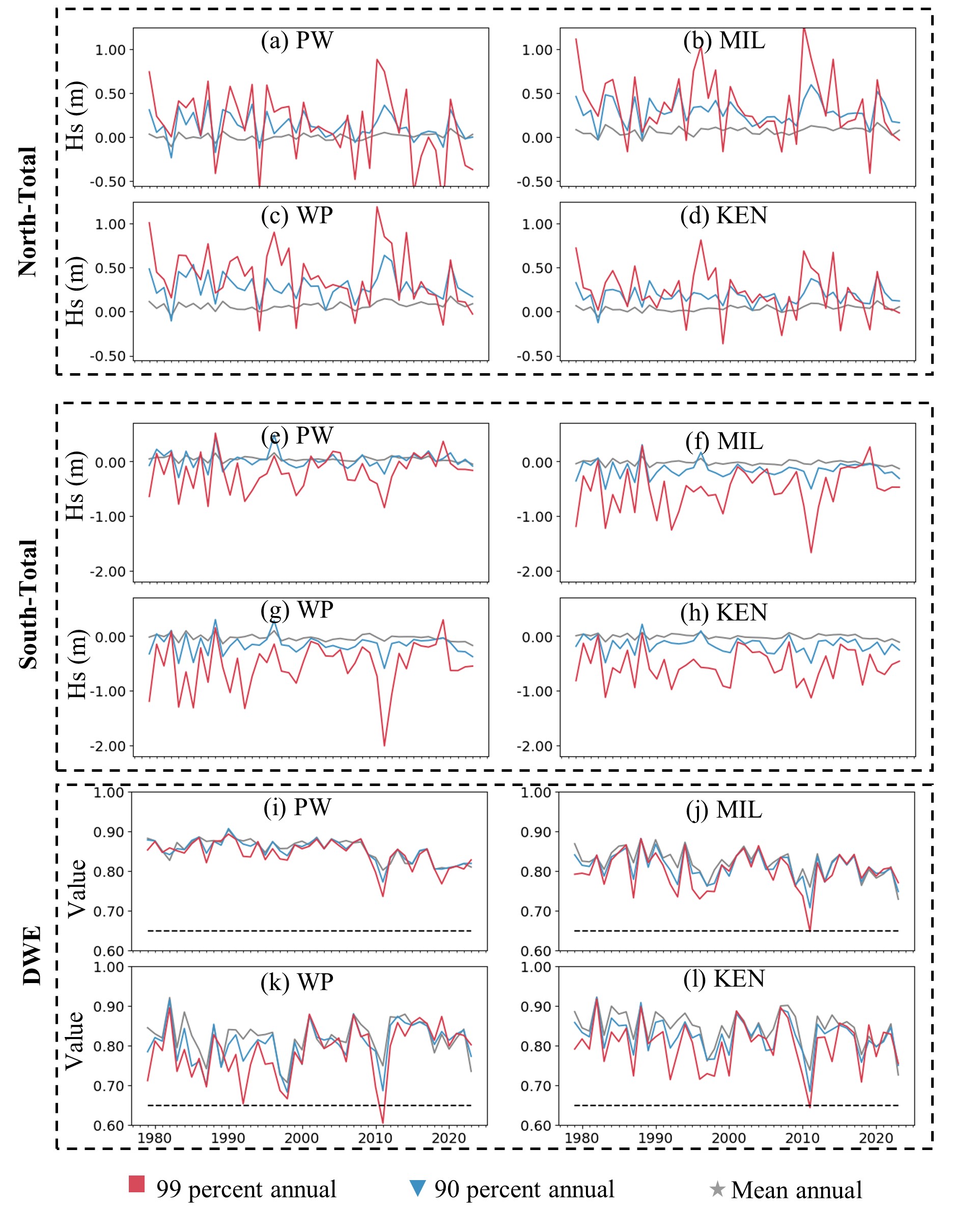}
  \caption{Inter-annual patterns of wave directionality. (a) – (d) are the inter-annual wave height difference between for northern wave and total wave at four major cities (PW-Port Washington, MH-Milwaukee Harbor, WP-Wind Point, KH-Kenosha Harbor), respectively, (e) – (h) are the inter-annual wave height differences between southern waves and total waves at four major cities, respectively, and (i) to (l) are the inter-annual directional wave entropy at four major cities, respectively. }
  \label{fig:fig3.5}
\end{figure}

The inter-annual variability of wave heights for the northern waves, southern
waves, and the values of DWE were analyzed in terms of the annual mean, 90\%
extreme values, and 99\% extreme values, as illustrated in Figure
\ref{fig:fig3.5}. During most of the years, northern waves exhibited greater
magnitudes than the southern waves, as shown in Figure \ref{fig:fig3.5}a-d.
However, during certain years (\eg 1992, 1998, and 2011), northern waves
declined. Differences were observed across the four locations, highlighting the
spatial variability in directional wave climate, where the northern sites (\eg
PW, MIL) experienced larger southern wave heights compared to the southern sites
(\eg KEN, WP). Figure \ref{fig:fig3.5}i-l demonstrates the annual variations
of DWE. During most of the years, the directionalities at the four locations
were bi-directional, while in some years (\eg 2011) the directionality changed
into uni-directionality at MIL, WP, and KEN. In addition, the comparison between
the mean, extreme 90\%, and extreme 99\% values for the northern and southern
wave heights and DWE was examined in Figure \ref{fig:fig3.5} . For both northern
and southern waves, the extreme 99\% wave heights (red lines in Figure
\ref{fig:fig3.5}) experienced significantly larger variations than the mean and
extreme 90\% wave heights (grey and blue lines in Figure \ref{fig:fig3.5}). For
instance, in 1987, the northern wave heights at the 99th percentile were the
largest group; however, in the subsequent year (1988), they decreased
significantly and became the lowest group. The huge variation in extreme groups
was also reflected in DWE (Figure \ref{fig:fig3.5}i-l), suggesting that
extreme waves may frequently shift directional dominance.

Furthermore, the inter-annual variability statistical metrics, including
p-values and slope estimated by statistical test, were summarized in Table
\ref{tab:tab3.1}. The annual trends in wave heights across different directional
groups exhibited notable inconsistencies. PW had a decreasing trend for northern
waves (-0.793, -5.900, and -18.729 mm/yr for mean, 90\%, and 99\%,
respectively), southern waves (-2.220, -2.082, and 0.764 mm/yr for mean, 90\%,
and 99\%, respectively), and decreasing trends for DWE (-1.354, -1.173, and
-1.114 for mean, 90\%, and 99\%, respectively). Nevertheless, not all subgroups
showed a significant linear trend (e.g., mean-north, 90\% and 99\% south). MIL
had trends as northern waves of 0.341, -1.869, and -11.423 mm/yr, southern waves
of -1.42, 0.321, and 1.755 mm/yr, and DWEs of -1.231, -0.821, and -0.341. Most
of them did not present a significant linear trend. WP had an increasing trend
at 0.467 mm/yr for northern mean waves but decreasing trends at -1.655 and
-13.362 mm/yr for northern 90\% and 99\%. The trends of southern wave heights in
WP were -1.814, 0.411, and 2.217 mm/yr. The DWE also had trends at -0.079,
0.959, and 1.988, respectively. Lastly, KEN showed a decreasing trend at 0.349,
0.326, and -7.075 mm/yr for northern mean, 90\%, and 99\%, while a consistent
decreasing trend for southern wave (-1.532, -1.183, and -1.158 mm/yr) and DWE
(-1.226, -0.795, and -0.006). Both WP and KEN present significant trends for
their southern mean waves. In conclusion, only a few directionality groups
exhibit statistically significant annual trends, and the significance observed
in the mean group does not necessarily imply significance in the extreme group.

\input{resources/table3-1.tex}

\subsection{Intra-annual patterns}
\label{Intra-annual patterns}

\begin{figure}[htbp]
  \centering
  \includegraphics[width=0.8\textwidth]{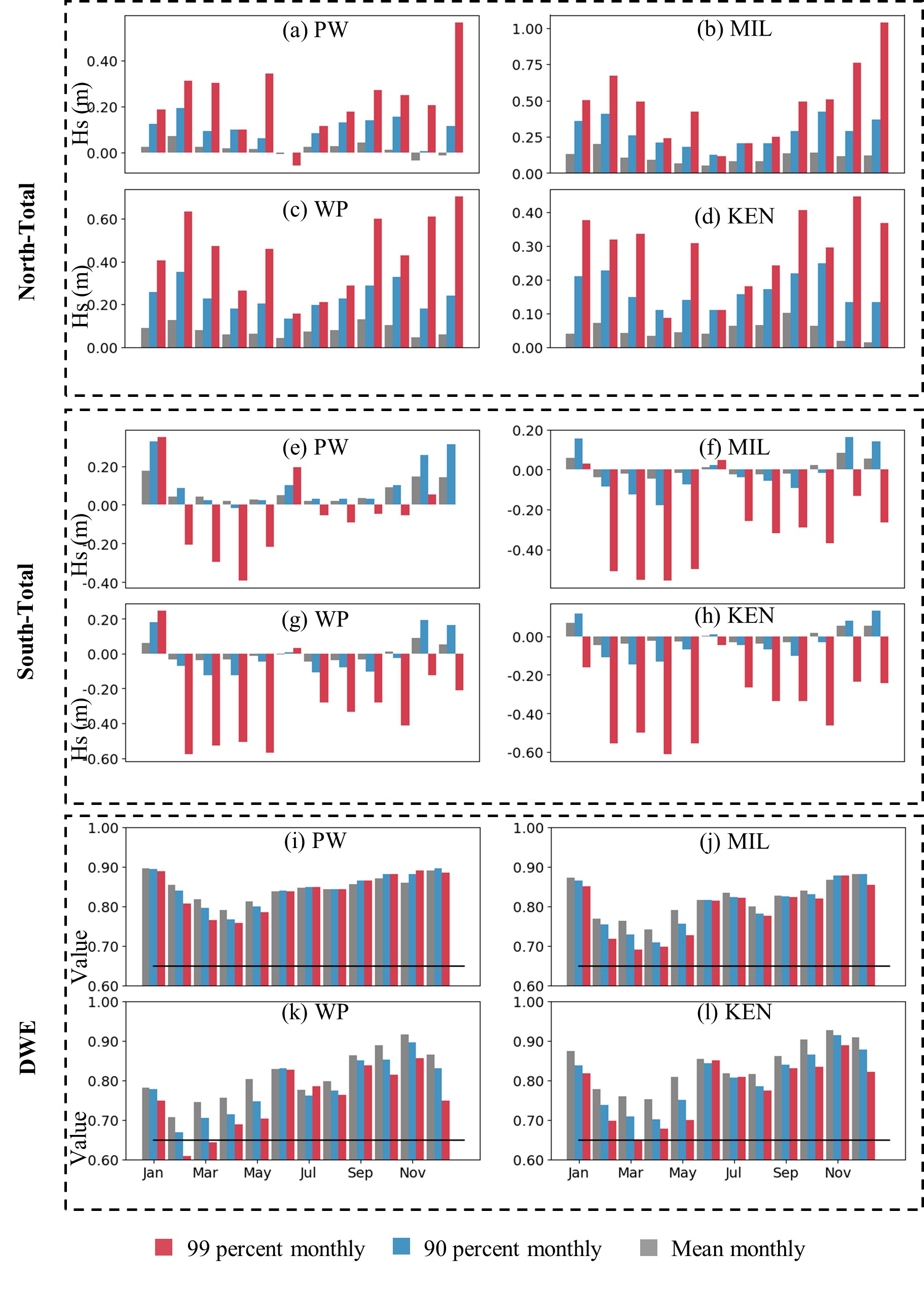}
  \caption{Intra-annual patterns of wave directionality. (a) – (d) are the intra-annual wave height differences between northern and total waves at four major cities (PW-Port Washington, MIL-Milwaukee Harbor, WP-Wind Point, KEN-Kenosha Harbor), respectively; (e) – (h) are the intra-annual wave height differences between southern and total waves at four major cities, respectively; and (i) to (l) are the intra-annual directional wave entropy at four major cities, respectively.}
  \label{fig:fig3.6}
\end{figure}

The intra-annual patterns of wave height and the Directional Wave Entropy (DWE)
were presented in Figure \ref{fig:fig3.6}. The northern wave was larger than the
southern wave in most months at most stations, except in June and November at PW
(Figure \ref{fig:fig3.6}a). Southern wave heights exhibited significant
seasonal variation, decreasing during the summer months and increasing during
winter. All three groups—mean, 90th percentile extreme, and 99th percentile
extreme—exhibited a similar monthly pattern, with northern wave heights
increasing during the winter months, particularly in December, January, and
February. This pattern flipped in the summer season such that the southern waves
became dominant, as shown in Figure \ref{fig:fig3.6}e-h. The DWE also showed a
clear seasonal pattern among the mean, 90\%, and 99\% extreme groups. During the
winter months, the DWEs of three groups for all four stations were approximately
0.886, 0.875, and 0.849, indicating a strong bi-directionality. Conversely,
during the spring season, the DWE decreased approximately 10\%, up to 0.802,
0.777, and 0.741, suggesting a weakening bi-directionality. In WP, DWE dropped
below the threshold during the spring season, showing that the dominant waves
shifted to southern waves at this location, leading to a strong
uni-directionality.  

\subsection{Spatial variabilities}
\label{Spatial variabilities}

\begin{figure}[htbp]
  \centering
  \includegraphics[width=0.8\textwidth]{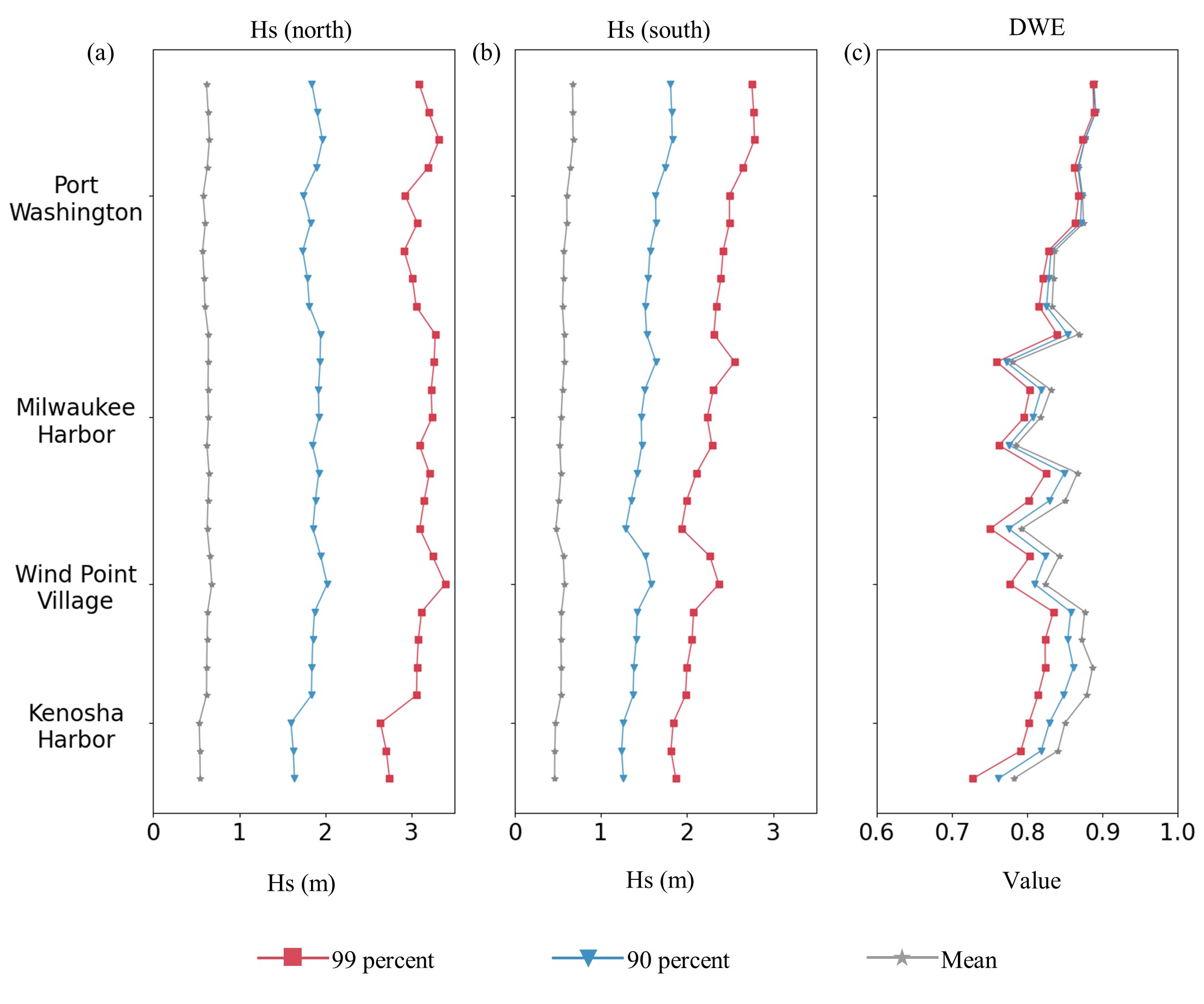}
  \caption{Spatial variability of wave directionality. (a) the spatial distribution of northern wave height at all WIS station sites, (b) the spatial distribution of southern wave height at all WIS station sites, and (c) the spatial distribution of DWE at all WIS station sites. The wave directionalities are computed for the mean, 90$\%$, and 99$\%$.}
  \label{fig:fig3.7}
\end{figure}

Spatial variability of wave characteristic statistics, including the mean and
extreme values (90\% and 99\%) for northern, southern, and DWE, for all 26 WIS
stations along the coastal regions of the four counties in Wisconsin, is
demonstrated in Figure \ref{fig:fig3.7}. For northern waves (Figure
\ref{fig:fig3.7}a), the mean, 90\%, and 99\% wave heights of all stations are
0.566, 1.602, and 2.677 meters, respectively. The wave heights exhibited a
gradual decreasing trend from the north to south for both mean and extreme
values. However, the decrease rate was not consistent, as it depends on the
specific stations and the mean or extreme groups. For the southern waves (Figure
\ref{fig:fig3.7}b), the mean, 90\%, and 99\% wave heights of all stations are
0.563, 1.572, and 2.452, respectively, which were slightly lower than those of
the northern waves. The southern wave directionality exhibited a higher
decreasing rate than the northern waves. The 99 percent extreme wave height also
declined at a higher rate than the mean and 90 percent. On the other hand, the
variability in DWE was more inconsistent across the mean and extreme values. The
average DWE values for three groups were 0.847, 0.835, and 0.817, indicating
obvious bi-directionality. The DWE decreased from the northern to southern
sites, indicating a decreasing bi-directionality as one moves southwards. This
spatial pattern aligned with the increased occurrence of northern waves at
southern sites. Notably, the values of 99\% DWE were smaller than 90 percent and
mean DWE, indicating that extreme wave conditions had stronger
uni-directionality. 

\subsection{Extreme wave climates}
\label{c3_Extreme wave climates}

\begin{figure}[htbp]
  \centering
  \includegraphics[width=0.8\textwidth]{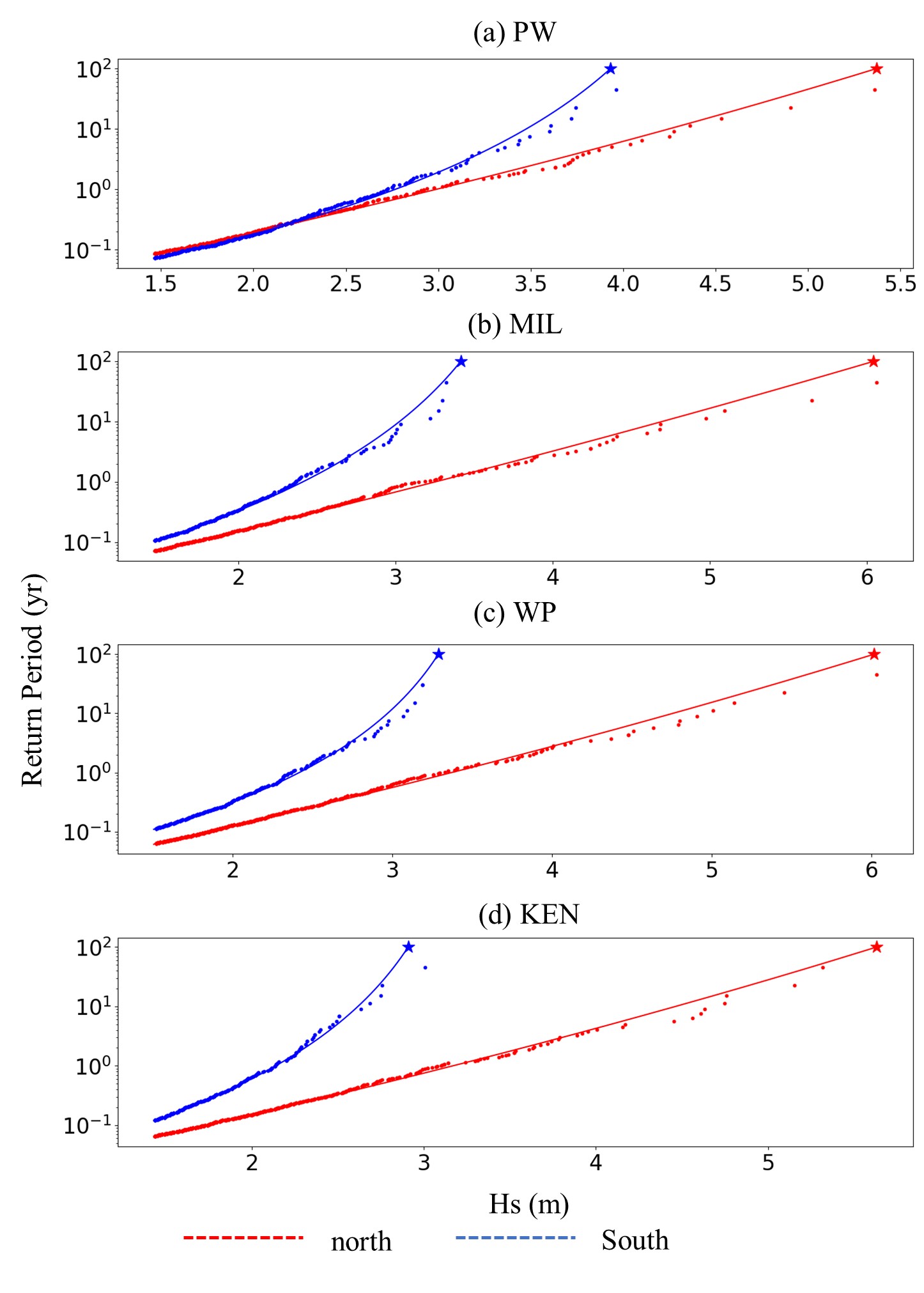}
  \caption{Return periods (years) of extreme wave height using POT for wave directionality. (a) - (d) show extreme wave directionality at four selected cities (PW-Port Washington, MIL-Milwaukee Harbor, WP-Wind Point, and KEN-Kenosha Harbor). The blue and red stars represent the 100-year return periods projected by the Generalized Pareto Distribution for northern and southern waves, respectively.}
  \label{fig:fig3.8}
\end{figure}

Figure \ref{fig:fig3.8} depicts the varied return period with corresponding
extreme wave heights, which were fitted and extrapolated by the Gumbel
distributions. For 1-year return periods, the estimated extreme wave heights
ranged approximately between 2 and 4 meters for four selected stations from 1979
to 2023, with minimal differences between dominant wave directions or station
locations. For the longer return period (100-year), the extreme wave heights
varied with different wave directionalities. The 100-year extreme wave heights
for southern and northern waves were as follows: 3.932 meters and 5.371 meters
in PW, 3.415 meters and 6.042 meters in MIL, 3.289 meters and 6.015 meters in
WP, and 2.909 meters and 5.633 meters in KEN. All sites exhibited a consistent
pattern, indicating that extreme wave heights were greater in the northern wave
direction, which has also been demonstrated in Figure \ref{fig:fig3.4}. The
existence of bi-directional extreme wave climate showed the potential
implications in the structure design. While designing a coastal structure, the
extreme waves might be underestimated without considering the wave
directionality.

\section{Discussions}
\label{c3_Discussions}

\subsection{Sensitivity of directionality characterization}
\label{c3_Sensitivity of directionality characterization}

\begin{figure}[htbp]
  \centering
  \includegraphics[width=0.8\textwidth]{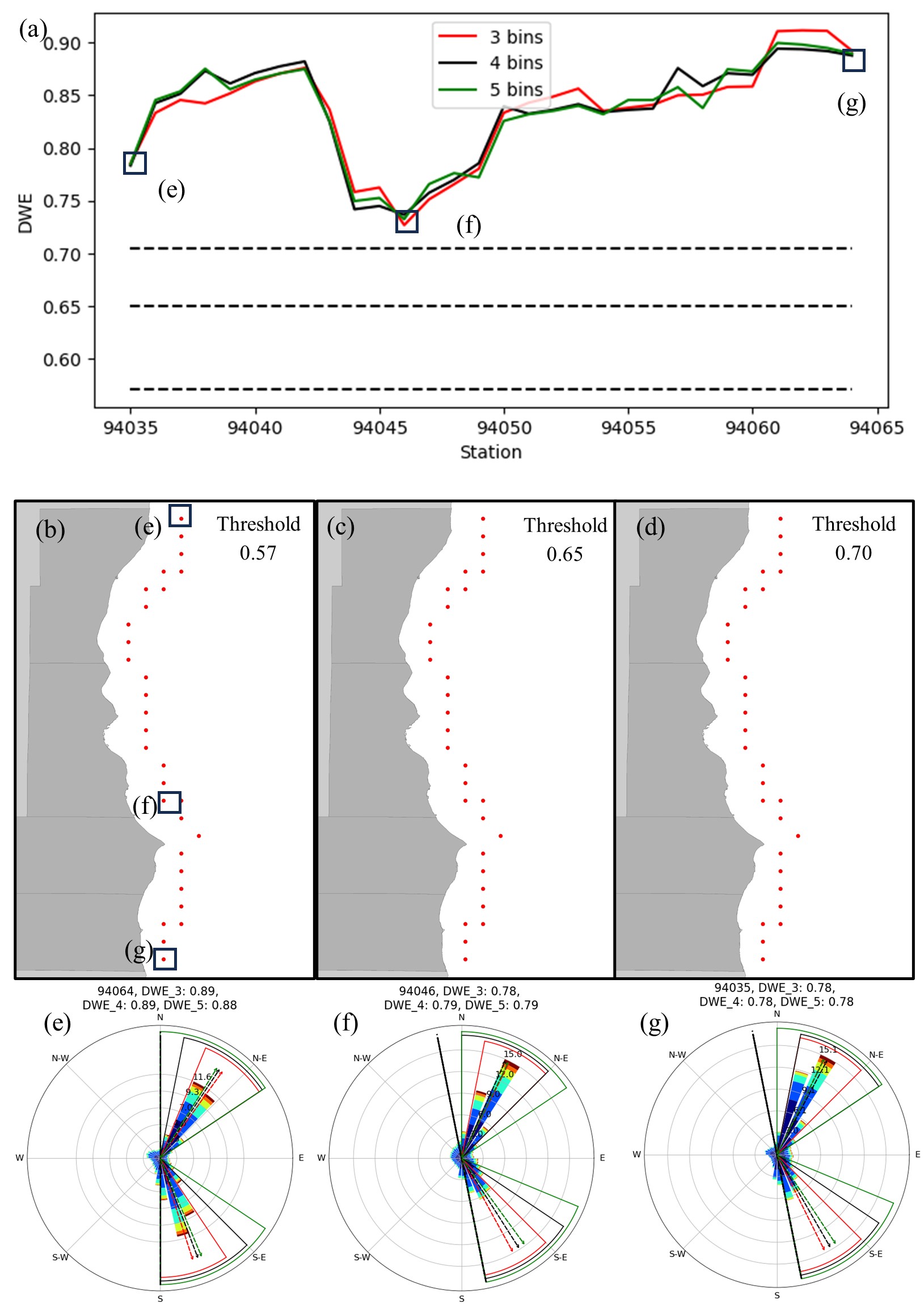}
  \caption{Wave directionality for different DWE thresholds and directional bins. (a) DWE values for all selected bins (red, black and green curves) and all thresholds (black dashed lines). Subplots (b), (c), and (d) are the maps of wave directionality (red dots are bi-directional and green dots are uni-directional) under three different thresholds: (b) 0.57, (c) 0.65, and (d) 0.70. Subplots (e), (f), and (g) are the wave roses at three representative locations under different directional bins. The reed frame is the 3-bins, black frame is 4-bins and green frame is 5-bins. The colored arrows are the averaged wave components for these directional bins, respectively}
  \label{fig:fig3.9}
\end{figure}

Two key variables influence the characterization of directionality: the number
of directional bins chosen and the DWE threshold value applied. Figure
\ref{fig:fig3.9} presents an analysis of how the number of directional bins and
DWE threshold values influence the characterization of wave directionality.
Three sets of directional bin criteria—3, 4, and 5 bins—and three threshold
levels—0.57, 0.65, and 0.70—were selected to classify several WIS stations
located in the southern regions of the study area (Figure \ref{fig:fig3.9}a).
The three threshold values were selected to represent common wave scenarios
involving two oblique wave systems (with a relative angle of $45^\circ$) and
significant wave height ratios of 3:1 (threshold 0.57), 2:1 (threshold 0.65),
and 1:1 (threshold 0.70), respectively. Across all stations within the study
area, wave directionality consistently exhibits a bi-directional pattern under
all tested criteria, as shown in Figure \ref{fig:fig3.9}b–d. Notably, panels
(b), (c), and (d) in Figure \ref{fig:fig3.9} illustrate the results using the
4-bin classification, as the choice of bin number has minimal impact on the
overall directionality characterization. This observation is further supported
by Figure \ref{fig:fig3.9}e–g, which displays wave roses at three selected
locations. The three different directional bin configurations—represented by the
colored frames in Figure \ref{fig:fig3.9}e–g—yield similar mean wave
components, as indicated by the three colored arrows. This suggests that the
number of directional bins has minimal influence on the DWE at these locations.
This can be attributed to the fact that dominant wave energy is typically
concentrated within a narrow angular range, resulting in negligible variation in
both the averaged wave directions (represented by the arrows in panels e–g) and
wave heights. Figure \ref{fig:fig3.2}a also presents the sensitivity of the
threshold for the whole lake. A low DWE threshold tends to overestimate
bi-directionality in Lake Michigan, whereas a high threshold may underrepresent
it. This is thus suggested for choosing an appropriate threshold when working
with other areas. The details of directionality in Lake Michigan will be
presented in the following discussion sections.

\subsection{Correlations of DWE with wave heights and wind speeds}
\label{c3_Correlations of DWE with wave heights and wind speeds}

\begin{figure}[htbp]
  \centering
  \includegraphics[width=0.8\textwidth]{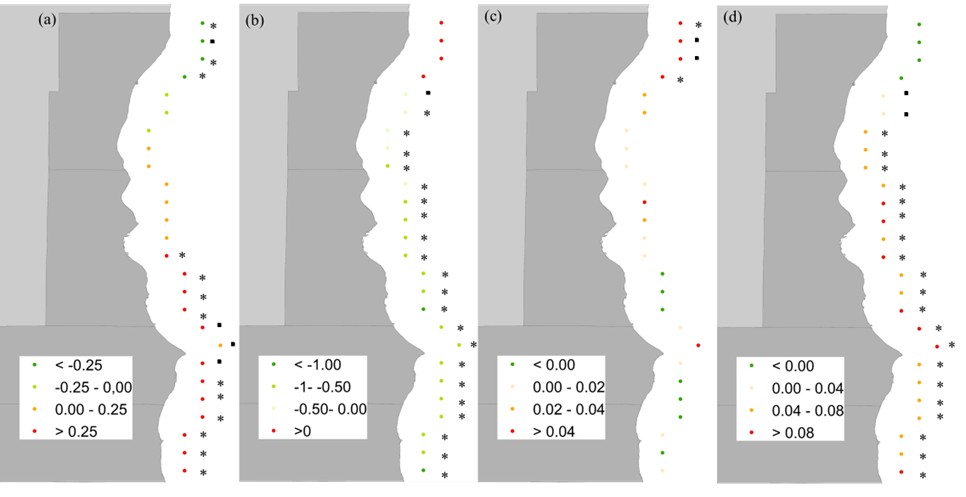}
  \caption{Linear regressions between DWE and other factors at 26 WIS stations. The correlated factors include (a) northern wave height, (b) southern wave height, (c) northern wind speed, and (d) southern wind speed. The colors of dots represent the slopes of regression. The black squares and asteroids represent the stations that have significant linear relationships with p-values less than 0.1 and 0.05, respectively.}
  \label{fig:fig3.10}
\end{figure}

To investigate the relationships of DWE (absolute values) with significant wave
height and wind speed, linear regression analyses were performed using data from
26 WIS stations. Monthly wave heights from northern (Figure
\ref{fig:fig3.10}a) and southern (Figure \ref{fig:fig3.10}b) directions, as
well as monthly wind speeds from northern (Figure \ref{fig:fig3.10}c) and
southern (Figure \ref{fig:fig3.10}d) directions, were respectively used to
calculate the slopes of regression and the p-values. Notably, the wind speed
data were extracted from WIS, which is the interpolated data from CFSR
\citep{saha_ncep_2010}. For northern wave heights in Figure
\ref{fig:fig3.10}a, negative correlations with DWE with strong confidence
levels (p-values < 0.05) were observed at stations located in Ozaukee County,
whereas positive trends occurred at southern stations in Milwaukee, Racine, and
Kenosha counties. Conversely, in Figure \ref{fig:fig3.10}b, southern wave
heights show negative linear trends across most WIS stations except for a few
stations in the northern side. The differences in correlation patterns with DWE
may be influenced by wind fetch length, as the wave heights are positively
correlated with the length of the fetch in Lake Michigan
\citep{mason_effective_2018}. With that, from southern station to northern
station, the increasing of southern wind fetch leads to an increasing southern
wave height, resulting in a decreasing correlation of DWE with both northern
waves and southern waves. Nevertheless, further examining the correlations
between DWE and wind speed from both north and south directions (in Figure
\ref{fig:fig3.10}c-d), it is found that there was no clear spatial pattern
observed in the slope of linear regression for either northern or southern wind
speeds. This observation highlights the need for future research to clarify the
relationship between wave directionality and wind climate.

\subsection{Wave directionality in Lake Michigan}
\label{Wave directionality in Lake Michigan}

\begin{figure}[htbp]
  \centering
  \includegraphics[width=0.8\textwidth]{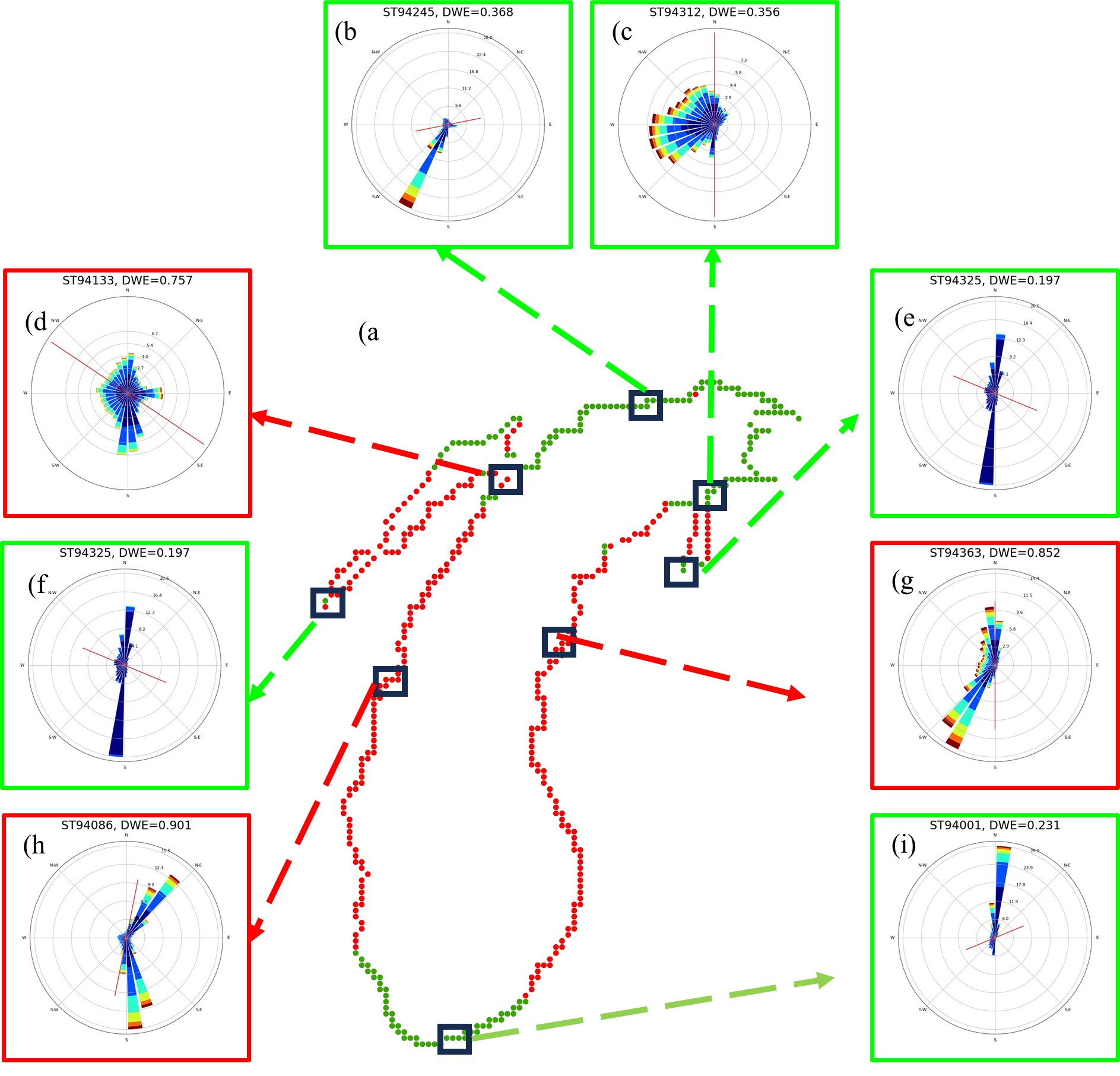}
  \caption{Directionality in Lake Michigan. (a) the wave directionalities of Lake Michigan in 490 WIS stations. The red dots are stations where the wave directionality is bi-directional, and the green dots are the locations where the wave directionality is uni-directional. (b) to (i) are the wave roses at several representative locations in Lake Michigan.}
  \label{fig:fig3.11}
\end{figure}

Characterization for directionalities of nearshore wave climate was extended to
the entire shoreline of Lake Michigan (Figure \ref{fig:fig3.11}). Based on wave
direction distribution extracted from 490 WIS stations around Lake Michigan,
there are two primary types of wave directionality in Lake Michigan:
uni-directional waves and bi-directional waves. The uni-directional wave
climates, in which waves are mostly distributed in a single direction, are most
likely to occur in the northern and southern shores of Lake Michigan (the red
dots in Figure \ref{fig:fig3.11}a). In northern Lake Michigan, waves tend to
come from the southwest (\eg Figure \ref{fig:fig3.11}b). Oppositely, the waves
will be most likely from the northern direction in southern Lake Michigan (\eg
Figure \ref{fig:fig3.11}i). The bi-directional wave climates, on the contrary,
are likely to occur at the western and eastern shorelines of Lake Michigan (the
green dots in Figure \ref{fig:fig3.11}). While western Lake Michigan receives
northeastern and southeastern waves (as shown in Figure \ref{fig:fig3.11}h),
eastern Lake Michigan receives the northwestern and southwestern waves (as shown
in Figure \ref{fig:fig3.11}g). Notably, in the Green Bay area (Figure
\ref{fig:fig3.11}f) and the West/East Arm of Grand Traverse Bay (Figure
\ref{fig:fig3.11}e), wave directionality was incorrectly classified as
uni-directional due to limitations in the DWE characterization. This issue will
be further examined in the following section. The different directionalities in
Lake Michigan result from different sizes of wind fetch fields. The shape of
Lake Michigan is north-south elongated with an aspect ratio of approximately 4,
which indicates that the fetch length in the northern-southern orientation is
significantly longer than the western-eastern orientation. In Lake Michigan,
most of the waves are affected by the wind fetch, so larger fetch lengths can
generate larger wave heights \citep{mason_effective_2018}. For example, the WIS
wave station 94001, which is located at the southern edge of Lake Michigan, is
most likely to receive the northern wave because of the 450-km fetch from north
to south. On the contrary, station 94086, which is located on the western shore
of Lake Michigan, tends to have bi-directional waves in both northeastern and
southeastern directions due to equal fetch fields existing in these directions. 

\subsection{Embayed beach morphology under wave directionality}
\label{Embayed beach morphology under wave directionality}

\begin{figure}[htbp]
  \centering
  \includegraphics[width=0.8\textwidth]{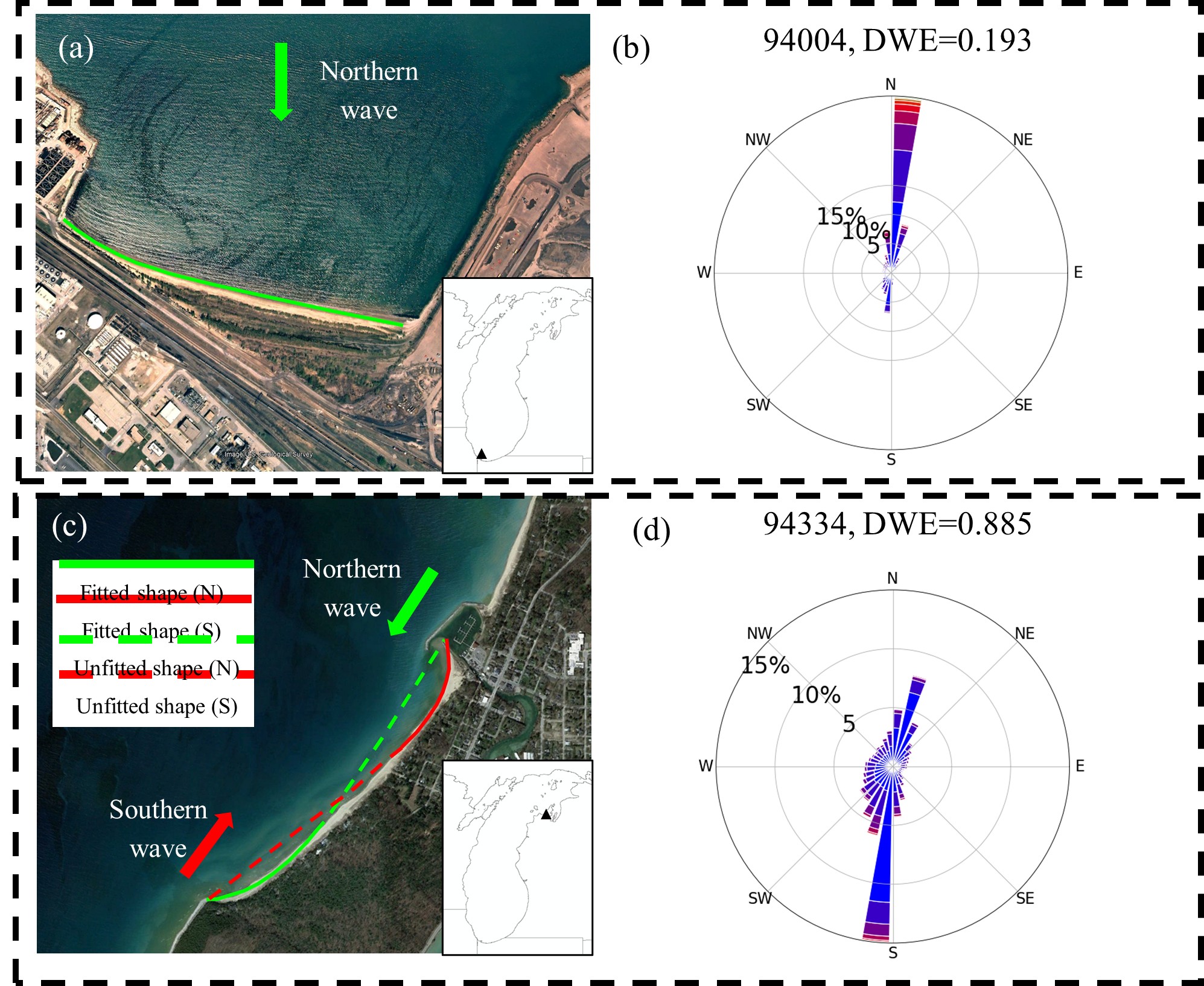}
  \caption{Embayed beaches shapes under different directionalities. (a) uni-directional wave climate at East Chicago Harbor, Illinois, associated with (b) its wave rose. (c) bi-directional wave climate at Le Land, Michigan, associated with (d) its wave rose. Beaches shapes were fitted by MepBay software using two wave diffraction points for bi-directional wave climate and one wave diffraction point for uni-directional wave climate.}
  \label{fig:fig3.12}
\end{figure}

One important implication of bi-directional wave climate is its impact on beach
morphology, especially embayed beaches
\citep{wiggins_coastal_2019,wiggins_regionally-coherent_2019}. Embayed beach is
a special landscape that is semi-enclosed between headlands and can be widely
seen in the urban and natural area of Lake Michigan \citep{mattheus_great_2022}.
Embayed beach shape can be approximated by a parabolic equation, which is
controlled by physical wave diffraction point at headland
\citep{moreno_equilibrium_1999}. Nevertheless, for a bi-directionality,
conventional beach shapes under a parabolic curve might not accurately depict
the headland bay beach shape. A common solution to this issue is to generate
multiple shapes with wave diffractions from bi-directionality, then combine the
shapes together. Figure \ref{fig:fig3.12} shows different beach shapes under
bi-directional wave and uni-directional wave climate, generated by MepBay
software \citep{da2003visual}. For beaches influenced by uni-directional wave
climates (Figure \ref{fig:fig3.12}a–b), shoreline morphology tends to be
asymmetrical, with beach width typically increasing near the edges of headlands
due to wave diffraction effects. Unlike uni-directionality, the shapes of
beaches for bi-directionality are relatively symmetrical curves (Figure
\ref{fig:fig3.12}c-d), which can be better approximated by two parabolic
curves. This indicates that the beach accretes at both ends of the bay under
bi-directional wave climates. Embayed beach morphology is often controlled by
nearshore sediment transport, especially longshore sediment transport. For a
semi-closed embayed beach, oblique waves drive the longshore sediment transport,
leading to erosion in the updrift beach and deposition in the downdrift beach
\citep{moreno_equilibrium_1999,loureiro_24_2020}. Bi-directional wave climates,
which indicate waves travel from two opposite directions, tend to drive
longshore sediment transport in a bi-directional pattern, leading to deposition
in both downdrift and updrift directions. To verify this, the longshore sediment
transport was estimated in the sites using the CREC equation, which is shown in
Figure \ref{fig:fig3.3}a. The bi-directional wave climate will lead to
bi-directional longshore sediment transport (\eg Figure \ref{fig:fig3.3}b)
more than the uni-directional wave climate (\eg Figure \ref{fig:fig3.3}a).
Hence, under bi-directional wave climate, beach morphology is present in a
symmetrical shape, which is fitted by two spiral curves. 

\subsection{Limitations of DWE and future applications}
\label{c3_Limitations of DWE and future applications}

DWE is based on the concept of cross-entropy and is simplified under the
assumption of a maximum of two dominant wave components on the water-facing side
of the shoreline. A primary limitation of this approach lies in its underlying
assumption, which is generally valid for most coastal regions but may not be
suitable for other environments such as offshore areas, bays, and straits. In
such areas, wave directionality is typically more complex—often exceeding
bi-directional patterns—and shoreline orientation is no longer a suitable
reference for capturing the full range of wave components. For example, in
Figure \ref{fig:fig3.11}e-f, the wave directionality was
mischaracterized because the second wave component was excluded due to its
alignment relative to the shoreline orientation. This suggests that filtering
wave incidents based on shoreline orientation may be ineffective in certain
locations, such as bays and straits, where wave angles are more complex.

These limitations stem from assumptions about the wave climate in Lake Michigan.
Excluding landward wave components when characterizing wave directionality is
generally reasonable, as such waves are typically of low magnitude. However,
with modifications, this approach could be extended to offshore areas of other
lakes and oceans. For example, the DWE equation can be modified by incorporating
a third entropy term to account for the presence of a third dominant direction.
Meanwhile, the characterization of wind directionality and its interrelationship
with wave directionality remain largely unexplored, highlighting an important
direction for future research.

\section{Conclusion}
\label{c3_Conclusion}

This study characterized the wave directionality and revealed the temporal and
spatial patterns in the southeastern Wisconsin coast. An innovative
index—Directional Wave Entropy (DWE)—was developed to characterize the
directionality of wave climate in this study. The results of DWE in four
selected WIS stations showed a bi-directional wave climate for historical wave
data from 1979 to 2023. The wave height for both northern and southern wave
directionality did not present a significant inter-annual trend for mean and
extreme wave conditions. The annual DWE shows a long-term bi-directionality
throughout 44 years. The intra-annual wave climate presented a clear seasonal
pattern for both northern and southern groups: the wave height increased during
the winter season. The monthly DWE also exhibited a seasonal pattern, with lower
values observed during the summer months, suggesting that stronger
bi-directionality tends to occur in the summer season. The extreme analysis of
wave directionality was also performed, showing that the extreme wave height for
the 100-year return period varied significantly between the northern and
southern wave directionality. Furthermore, this study discussed the impact of
selecting the DWE threshold on the characterization of wave directionality and
found that a low threshold can underestimate the bi-directionality. This study
also explored the correlation between DWE with wave height and wind speed,
respectively. The result showed a correlation between wave index and wave
climate but not wind climate. In addition, the wave directionality along the
entire Lake Michigan coast was discussed, which reveals the presence of
bi-directional and uni-directional wave climate. The presence of
bi-directionality and uni-directionality follows a specific spatial distribution
which is closely related to the topological shape of Lake Michigan. Furthermore,
the impact of wave directionality on headland bay beach morphology was also
examined at two beaches as examples. The difference of beach morphology in the
two sites suggested such an implication that wave bi-directionality tends to
cause symmetrical beach morphology. Lastly, the limitations and potential of DWE
approaches were discussed, indicating needs for future studies.

\bibliographystyle{unsrt}  
\bibliography{ref}

\end{document}

%% file: resources/table3-1.tex
\begin{table}[htbp]
\centering
\caption{P value and Slope for Northern wave, Southern wave, and DWE at different confidence levels}
\begin{tabular}{|c|c|cc|cc|cc|}
\hline
\textbf{Sites} &  \textbf{Groups}     & \multicolumn{2}{c|}{\textbf{Northern wave}} & \multicolumn{2}{c|}{\textbf{Southern wave}} & \multicolumn{2}{c|}{\textbf{DWE}} \\
\cline{3-8}
               &       & \textbf{P value} & \textbf{Slope} & \textbf{P value} & \textbf{Slope} & \textbf{P value} & \textbf{Slope} \\
\hline
\multirow{3}{*}{PW}  & Mean & 0.091 & -0.793 & \textbf{0.001} & -2.220 & \textbf{0.000} & -1.354 \\
                     & 90\% & \textbf{0.039} & -5.900 & 0.333 & -2.082 & \textbf{0.000} & -1.173 \\
                     & 99\% & \textbf{0.006} & -18.729 & 0.762 & 0.764 & \textbf{0.002} & -1.114 \\
\hline
\multirow{3}{*}{MIL} & Mean & 0.591 & 0.341 & \textbf{0.026} & -1.420 & \textbf{0.003} & -1.231 \\
                     & 90\% & 0.525 & -1.869 & 0.762 & 0.321 & 0.062 & -0.821 \\
                     & 99\% & 0.102 & -11.423 & 0.512 & 1.755 & 0.451 & -0.341 \\
\hline
\multirow{3}{*}{WP}  & Mean & 0.512 & 0.467 & \textbf{0.045} & -1.814 & 0.899 & -0.079 \\
                     & 90\% & 0.618 & -1.655 & 0.899 & 0.411 & \textbf{0.094} & 0.959 \\
                     & 99\% & \textbf{0.056} & -13.362 & 0.500 & 2.217 & 0.005 & 1.988 \\
\hline
\multirow{3}{*}{KEN} & Mean & 0.525 & 0.349 & \textbf{0.035} & -1.532 & \textbf{0.015} & -1.226 \\
                     & 90\% & 0.822 & 0.326 & 0.395 & -1.183 & 0.083 & -0.795 \\
                     & 99\% & 0.168 & -7.075 & 0.604 & -1.158 & 0.992 & -0.006 \\
\hline
\end{tabular}
\label{tab:tab3.1}
\end{table}